\documentclass[pre,preprint,showpacs]{revtex4}

\usepackage{graphicx}

\usepackage{amsmath,amssymb}
\usepackage{bm}
\usepackage{cancel}

\usepackage{color}

\begin{document}
\title{Stochastic Model of Tumor-induced Angiogenesis: Ensemble Averages and Deterministic Equations}
\author{ F.\ Terragni$^1$, M.\ Carretero$^1$, %M.\ Alvaro$^1$, 
V.\ Capasso$^{2}$ and L.\ L.\ Bonilla$^1$}
\affiliation {$^1$G. Mill\'an Institute, Fluid Dynamics, Nanoscience
and Industrial Mathematics, Universidad Carlos III de Madrid, 28911
Legan\'es, Spain} \affiliation{$^2$ ADAMSS, Universit\'a degli Studi
di Milano, 20133 MILANO, Italy}

\date{\today}
\begin{abstract} 
A recent conceptual model of tumor-driven angiogenesis including branching, elongation, and anastomosis of blood vessels captures some of the intrinsic multiscale structures of this complex system, yet allowing to extract a deterministic integro-partial differential description of the vessel tip density [Phys. Rev. E {\bf 90}, 062716 (2014)]. Here we solve the stochastic model, show that ensemble averages over many realizations correspond to the deterministic equations, and fit the anastomosis rate coefficient so that the total number of vessel tips evolves similarly in the deterministic and ensemble averaged stochastic descriptions. 
\end{abstract}

\maketitle

\section{Introduction}
\label{sec:intro}

Tumor growth in living tissues involves fast proliferating cells that need oxygen and nutrients. The latter are transported by vascular blood and, therefore, the vasculature about a growing tumor has to be substantially increased by angiogenesis, i.e., by creating new blood vessels from existing ones \cite{CJ2011,carmeliet_2005}. Angiogenesis is also essential for normal organ growth and repair \cite{GG2005,fruttiger}. The growth of blood vessel and of nerve networks presents common mechanisms that are fascinating to explore \cite{CT2005}. In recent years, understanding of the molecular mechanisms of angiogenesis has increased at an explosive rate and has led to the approval of anti-angiogenic drugs for cancer and eye diseases \cite{CJ2011}. In combination with experiments, mathematical and computational models of angiogenesis are an important part of these efforts \cite{sto91,morale_chaplain_1998,tong,lev01,morale:plank_sleeman:03,othmer,morale:sun_wheeler:05,morale:sun_wheeler_SIAM:05,morale:chaplian_2006,bau07,VK_morale_jomb,jac10,das10,swanson2011,preziosi,dejana14,bentley14,cotter,bon14,hec15}. 

Angiogenic systems are intrinsically complex multi-scale systems that present variations depending on whether they are associated to tumor or normal organ growth. A brief description of angiogenesis adapted from \cite{hec15} gives an idea of the disparity of scales involved in the process. Angiogenesis appears as a response to lack of oxygen (hypoxia). Hypoxic cells secrete vessel endothelial growth factor (VEGF). The growth factor diffuses through the surrounding extracellular matrix (ECM), but it is of course more abundant near the hypoxic cells. The endothelial cells (ECs) of a nearby blood vessel undergo phenotypic changes when reached by the VEGF and may get transformed into moving tip cells that start an angiogenic sprout. Tip cells do not proliferate. They move chemotactically toward the direction of increasing VEGF gradient secreting ECM degrading enzymes to progress. Notch signaling impedes neighboring ECs to become tip cells. Instead, they become stalk cells, proliferate, migrate and help building the capillary that was initiated by the tip cells. When tip and stalk cells migrate, they interchange types as the angiogenic sprout advances \cite{bentley14}. Vascular lumen forms and blood flows through the sprouting vessel. When a moving sprout meets another vessel, the tip cells can merge in a process called anastomosis. Anastomosis favors blood circulation which contributes to oxygenate the tissue and leads to a decrease in VEGF expression levels. Then the newly formed vessels become mature and the ECs turn quiescent. The length scales involved in angiogenesis range from sub-cellular (submicron) to macroscopic (millimeters). Sprouts advance a few millimeters per day. 

Angiogenesis models range from simple tip endothelial cell migration models (that do not describe the cellular scale) \cite{sto91,morale_chaplain_1998,tong,lev01,morale:plank_sleeman:03,othmer,morale:sun_wheeler:05,morale:sun_wheeler_SIAM:05,morale:chaplian_2006,VK_morale_jomb,cotter,bon14,mil08}, to stalk-tip cell based models (that distinguish between tip and stalk ECs and include other processes as proliferation, maturation and apoptosis) \cite{swanson2011,jac10,das10}, and to models capturing cell dynamics at cellular scale \cite{bau07,preziosi,bentley14}. A discussion of the different types of models and the current state of the art can be found in \cite{hec15}. Numerical solutions of detailed multiscale models combined with experiments have clarified important aspects of angiogenesis e.g. the interchange between tip and stalk ECs \cite{bentley14}. However, the complexity of these models makes their analysis quite difficult and the values of the parameters involved in complex models may be quite uncertain. 

For macroscopic lengths, older angiogenesis models postulate systems of coupled reaction-diffusion equations (RDEs) for VEGF concentrations, cell densities, etc \cite{lio77,cha93,cha95}. Later models typically treat in detail processes at some scales and coarse-grain over smaller scales. Migrating tip cell models consider tip ECs in a capillary sprout as particles and track their position, which means that the typical {\em mesoscopic} length scales for tip motion in these models are much larger than cell size ($\mu$m) but much smaller than macroscopic lengths (mm). In these models, the stalk cells in a growing vessel build the capillary following the wake of the cells at the vessel tip \cite{CJ2011}. Thus the idealized sprout comprises the present and all previous positions of the vessel tip. The motion of the tip cells is a stochastic process consistent with some continuum partial differential equation (PDE) for vessel tip density \cite{morale_chaplain_1998}, or with some master equation of a reinforced random walk \cite{othmer,morale:plank_sleeman:03}, and rules for random branching of tips and for anastomosis. Alternatively, deterministic rules for branching and motion can be set for tips moving on a spatially random ECM \cite{morale:sun_wheeler:05,morale:sun_wheeler_SIAM:05,mil08}. Tip motion is coupled to RDEs for continuum fields such as VEGF, matrix degrading enzymes, fibronectin, etc \cite{morale_chaplain_1998,tong,othmer,morale:plank_sleeman:03}. In these tip cell models, vessel tip  densities are calculated numerically and no evolution equations for them have been derived (continuum tip density PDEs \cite{morale_chaplain_1998} or master equations for reinforced random walks \cite{othmer,morale:plank_sleeman:03} considered in the respective models do not contain tip branching and anastomosis). Evolution equations for vessel tip densities are important because they give an alternative deterministic description of angiogenesis at mesoscopic lengths. In turn, such deterministic description may be amenable to  analyses of stability, long time behavior and control of solutions that could supplement numerical simulations of the models.

The program of deriving a deterministic description for the vessel tip density is typical of nonequilibrium statistical mechanics and it requires considering simple conceptual models of angiogenesis at first. Successful completion in simple cases may provide a template on how to carry out this program for more elaborate and realistic models \cite{hec15}. Angiogenesis models describing sub-cellular scales can be considered to be ``ab initio'' (similar to molecular dynamics), mesoscopic models are akin to kinetic theory (Boltzmann or Fokker-Planck equations), whereas macroscopic models are akin to continuum mechanics. Establishing connections between these levels of description is not straightforward, as the number of particles (vessel tips) involved in angiogenesis is rather modest. 

A particularly simple model focuses on the stochastic processes of branching, growth and vessel fusion (anastomosis) of vessel tips, driven by a single chemotactic field \cite{VK_morale_jomb,bon14}. Tip branching is a birth process because a new tip is created by branching, whereas anastomosis is a death process, as it occurs when a moving vessel tip finds an existing vessel, then merges with it and ceases to be actively moving. From the stochastic description, it is possible to derive a mean-field deterministic integrodifferential equation for the density of vessel tips coupled with a RDE for a tumor angiogenic factor (TAF) which acts as the chemotactic field \cite{bon14}. Although a tumor secretes different growth factors (vessel endothelial, fibroblast, platelet-derived and other growth factors) to attract blood vessels, we simplify the model by considering a single RDE for a generic TAF \cite{morale_chaplain_1998}. Appropriate boundary and initial conditions for the deterministic equations have also been established in \cite{bon14}. Other continuum fields such as fibronectin and ECM degrading enzymes useful to describe haptotaxis can be added to the model \cite{VK_morale_jomb} but will not be considered here for the sake of simplicity. The vessel tip density obeys a Fokker-Planck type equation with source terms corresponding to tip creation (branching) and annihilation (anastomosis, which is nonlocal in time). The latter term contains a rate constant that has to be calculated by comparison to the stochastic description. One of the motivations of this paper is to find the rate constant by comparing numerical simulations of the stochastic equations to numerical solutions of the deterministic integrodifferential equations found in \cite{bon14}. It turns out that the {\em same} deterministic equations hold for tip density and TAF fields that are {\em ensemble averages} over stochastic quantities. For the modest number of tips generated by simulations of our stochastic model, the law of large numbers that follows from the propagation of molecular chaos assumption is inapplicable {\em to a single replica}, and therefore the deterministic equations do not follow from it, as it had been conjectured previously \cite{VK_morale_jomb,bon14}. 

Mean field equations that follow from the law of large numbers are quite convenient as they hold for any given realization of the underlying {\em self-averaging} stochastic processes. If the fluctuations do not decay as the system scale increases, a deterministic description is still possible for averages over a sufficiently large number of realizations of the stochastic processes, i.e., within confidence bands for ensemble averages. With the deterministic interpretation of angiogenesis based on ensemble averages, the integral death term representing anastomosis has a natural meaning as being proportional to the occupation time density of a small volume in phase space (position and velocity of tips). While anastomosis is a history dependent process for a given replica of the stochastic process, it depends only on phase space for the ensemble of all possible independent realizations. 

The anastomosis term appearing in the deterministic integrodifferential equation for the vessel tip density is nonlocal in time and it points to a deficiency in the usual macroscopic descriptions of angiogenesis. In the latter, anastomosis is included as a local term that follows the usual mass action law (see \cite{cha93,cha95}, and references cited therein). Nonlocal anastomosis terms are likely to appear when equations for the vessel tip density are obtained from other migrating tip cell models, even from those obtaining rules for tip motion from macroscopic PDEs for EC densities that do not contain source terms \cite{morale_chaplain_1998}, or from those postulating a reinforced random walk consistent with a master equation plus branching and anastomosis rules \cite{othmer,morale:plank_sleeman:03}. 

The rest of the paper is as follows. Section \ref{sec:model} briefly summarizes the stochastic model of Ref. \cite{bon14}. Section \ref{sec:densities} explains how to extract tip and flux densities from ensemble averages of the stochastic processes. We derive an equation of Fokker-Planck type for the density of vessel tips and the TAF RDE in Section \ref{sec:deterministic}. This section also includes a discussion of the appropriate boundary and initial conditions. Numerical results for the nondimensional version of these equations and a calculation of the anastomosis coefficient by a fit to ensemble averages of the stochastic process are reported in Section \ref{sec:numerical} whereas section \ref{sec:conclusions} contains our conclusions. 

\section{Stochastic model} \label{sec:model}
As explained in Section \ref{sec:intro} and in Ref. \cite{bon14}, our stochastic model consists of a system of Langevin equations for the extension of vessel tips, a tip branching process and anastomosis or destruction of tips when they merge with existing vessels. In addition, we have diffusion of TAF and its consumption by advancing tips \cite{bon14}. Thus the stochastic model consists of
\begin{itemize}
\item {\em Vessel extension:}
\begin{eqnarray} 
d\mathbf{X}^i(t)&=&\mathbf{v}^i(t)\, dt,\nonumber\\
d\mathbf{v}^i(t)&=& \left[- k\, \mathbf{v}^i(t)+
\mathbf{F}\!\left(C(t,\mathbf{X}^i(t))\right)\!
\right]\! dt + \sigma\, d\mathbf{W}^i(t) \label{eq1}
\end{eqnarray}
(for $T^i<t<\Theta^i,$ where $T^i$ and $\Theta^i$ denote the random times of branching and of death for the $i$th tip, respectively). Here $\mathbf{X}^i(t)$ and $\mathbf{v}^i(t)$ are the position and the velocity of tip $i$ at time $t$, $\mathbf{W}^i(t)$ are i.i.d. standard Brownian motions, $C(t,\mathbf{x})$ is the TAF concentration, and $k$ (friction coefficient) and $\sigma$ are positive parameters. The chemotactic force is modelled as
\begin{eqnarray}
\mathbf{F}(C)&=& \frac{d_1}{(1+\gamma_1C)^q}\nabla_x C,
\label{eq2}
\end{eqnarray}
where $d_1$, $\gamma_1$, and $q$ are positive parameters.
\item {\em TAF diffusion and degradation:}
\begin{eqnarray} 
\frac{\partial}{\partial t}C(t,\mathbf{x})\!&\!=\!& \! d_2 \Delta_x C(t,\mathbf{x})%\nonumber \\ \!&\!-\!&\! 
-\eta C(t,\mathbf{x})\!\left| \sum_{i=1}^{N(t)}
\mathbf{v}^i(t)\delta_{\sigma_x}(\mathbf{x}-\mathbf{X}^i(t))\right|\!.\label{eq3}
\end{eqnarray}
Here $N(t)$ is the number of active tips at time $t$, $d_2$ (diffusivity) and $\eta$ are positive parameters, whereas $\delta_{\sigma_x}(\mathbf{x})$ is a regularized smooth delta function (e.g., a Gaussian with variances $l_x^2$ and $l_y^2$ proportional to $\sigma_x^2$ along the $x$ and $y$ directions, respectively) that becomes $\delta(\mathbf{x})$ in the limit as $\sigma_x\to 0$. In this limit (see Theorem 4 in page 489 of \cite{roussas}), the mean field term in this equation becomes the modulus of the tip flux. The sink term in \eqref{eq3} indicates that TAF is consumed in the process of enlarging the capillary: At the interval $dt$, tip $i$ advances to $\mathbf{v}^i(t)\, dt$, thereby enlarging the capillary by that vector, and TAF consumption should thus be proportional to $C$ times the tip flux modulus. Alternatively, a sink term proportional to $C$ times $\sum_{i=1}^{N(t)}|\mathbf{v}^i(t)|\,\delta_{\sigma_x}(\mathbf{x}-\mathbf{X}^i(t))$ could be used. The region around a vessel tip that affects TAF should be of the same order as the tip size that comprises about 10 cells \cite{bentley14}. The model considers mesoscopic length scales that are much larger than cell size ($\mu$m) but much smaller than macroscopic length (mm). For these mesoscopic lengths, the region about the tips affecting TAF is infinitesimal, so that the sink term in \eqref{eq3} can be considered local in space (in the limit as $\sigma_x\to 0$).
\item {\em Tip branching:} A tip $i$ is born at a random time $T^i$ from a moving tip (we ignore branching from mature vessels) and disappears at a later random time $\Theta^i$, either by reaching the tumor or by anastomosis. At time $T^i$, the velocity of the newly created tip $i$ is selected out of a normal distribution,
\begin{eqnarray}
\delta_{\sigma_v}(\mathbf{v})= \frac{e^{-|\mathbf{v}|^2/\sigma_v^2}}{\pi\sigma_v^2},\label{eq4}
\end{eqnarray}
with mean $\mathbf{v}_0$ and a narrow variance $\sigma_v^2$. In addition,  the probability that a tip branches from one of the existing ones during an infinitesimal time interval $(t, t + dt]$ is taken proportional to $\sum_{i=1}^{N(t)}\alpha(C(t,\mathbf{X}^i(t)))dt$, where
\begin{eqnarray}
\alpha(C)=\alpha_1\frac{C}{C_R+C}.\label{eq5}
\end{eqnarray}
Here $\alpha_1$ and $C_R$ (reference concentration) are positive parameters. Branching probability increases with increasing TAF density near the tumor following a saturating Holling type II or Michaelis-Menten law \cite{VK_morale_jomb}, which is consistent with experiments \cite{sholley}. Then the probability that a tip branches from one of the existing ones with velocity normally distributed about $\mathbf{v}_0$ during an infinitesimal time interval $(t, t + dt]$ is  $\sum_{i=1}^{N(t)}\alpha(C(t,\mathbf{X}^i(t)))\delta_{\sigma_v}(\mathbf{v}^i(t)-\mathbf{v}_0)dt$. The change per unit time of the number of tips in boxes $d\mathbf{x}$ and $d\mathbf{v}$ about $\mathbf{x}$ and $\mathbf{v}$ is
\begin{eqnarray}\nonumber
&&\sum_{i=1}^{N(t)}\alpha(C(t,\mathbf{X}^i(t)))\, \delta_{\sigma_v}(\mathbf{v}^i(t)-\mathbf{v}_0)\\
&&=\int_{d\mathbf{x}}\int_{d\mathbf{v}}\alpha(C(t,\mathbf{x)}) \delta_{\sigma_v}(\mathbf{v}-\mathbf{v}_0)%\nonumber \\&&\times 
\sum_{i=1}^{N(t)}\delta(\mathbf{x}-\mathbf{X}^i(t))\delta(\mathbf{v}-\mathbf{v}^i(t)) d\mathbf{x} d\mathbf{v}.
\label{eq6}
\end{eqnarray}

\item {\em Anastomosis:} When a tip meets an existing vessel, it joins it at that point and time, stops moving, and we cease counting it. This {\em death} process is called tip-vessel {\em anastomosis}.
\end{itemize}
Haptotaxis or motion towards a gradient of cellular adhesion sites present in the extracellular matrix can be treated as in \cite{VK_morale_jomb} but we do not include it here for simplicity. Vessel retraction and blood circulation in the vessels are also ignored. 

Our stochastic model is thus described by a set of Ito stochastic differential equations (SDEs), a marked point process describing tip branching (a birth process), and  a marked point process describing anastomosis (a death process). The latter process depends on the past history of a given realization of the overall stochastic process. Moreover, the TAF concentration is itself a random process since it depends on the stochastic evolution of the tips as indicated by Equation (\ref{eq3}).

\section{Tip density and flux given by ensemble averages}
\label{sec:densities}
To solve the Ito stochastic differential equations of the model (\ref{eq1})-(\ref{eq2}), we have used a standard stochastic Euler-Maruyama method \cite{kloeden,gardiner10} with time step $dt=0.003$. At each time step $dt$ and for each tip $i$, we extract a random number $U$ with equal probability between 0 and 0.4. A new tip branches out from $i$ at $\mathbf{x}=\mathbf{X}^i(t)$ only if $U<\alpha(C(t,\mathbf{X}^i(t)))\,dt/\tilde{v}_0^2$ (where $\tilde{v}_0=40\,\mu$m/hr is a typical velocity scale; see Section \ref{sec:numerical}). Its initial position is $\mathbf{x}$ and its initial velocity is selected out of a normal distribution with mean $\mathbf{v}_0$ and variance $\sigma_{v}^2$. 

\begin{figure}[ht]
\begin{center}
\includegraphics[width=11cm]{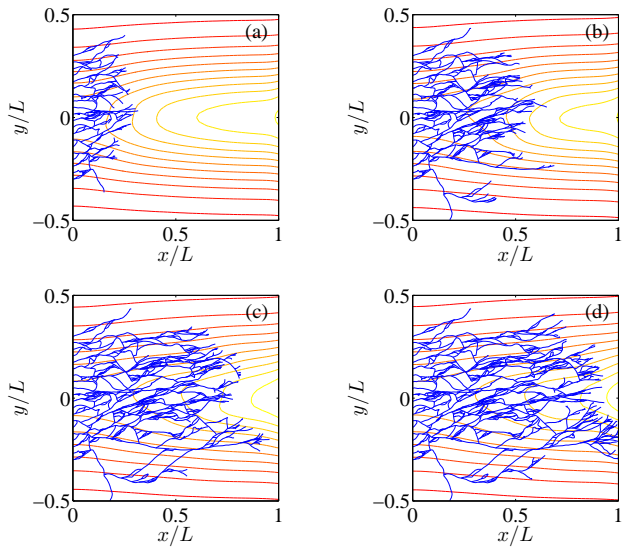} \qquad
\end{center}
\caption{(Color online) Snapshots of the vessel network inside a central square of side $L$ at times: (a) 12 h (46 active tips), (b) 24 h (60 active tips), (c) 32 h (78 active tips), and (d) 36 h (76 active tips). The level curves of the TAF density $C(t,\mathbf{x})$ are also depicted.  
\label{fig1}}
\end{figure}

A typical outcome of the simulations up to the first time of arrival to the tumor is depicted in Figure \ref{fig1}. Tips proliferate by branching but they tend to crowd in a relatively narrow region due to chemotaxis. Then anastomosis eliminates many vessel tips and, as a result, there never are enough tips for the law of large numbers to apply. Fluctuations (e.g., of the velocity) do not decay for the largest number of tips appearing in the simulations. Thus the stochastic model is not self-averaging and we cannot expect average quantities to be the same as those of a typical realization of the process for large enough number of tips. The vessel network may be quite different for different replicas of the stochastic process. However, the tip density defined below remains unaltered when the ensemble average is taken over a sufficiently large number of replicas and therefore we may expect a deterministic description of ensemble averaged densities. We shall first explain how the vessel tip density may follow from ensemble averages and then comment the results of simulations of the angiogenic stochastic process.

Let us consider a number $\mathcal{N}$ of independent replicas (realizations) of the angiogenic process with random initial conditions except that they all have the same initial number of vessel tips. For any replica $\omega$ at time $t$, we define the stochastic distribution of tips per unit volume in the $(\mathbf{x},\mathbf{v})$ phase space by
\begin{equation}
\!Q_N^*(t,\mathbf{x},\mathbf{v},\omega)\!\!=\!\!\!\sum_{i=1}^{N(t,\omega)}\!\!\!\delta_{\sigma_x}\!(\mathbf{x}-\mathbf{X}^i(t,\omega))\delta_{\sigma_v}\!(\mathbf{v}-\mathbf{v}^i(t,\omega)). \label{eq7}
\end{equation}
Here $\delta_{\sigma_v}(\mathbf{v})$ is given by (\ref{eq4}) and $\delta_{\sigma_x}(\mathbf{x})$ is the same type of Gaussian kernel of variance $\sigma_x^2$ that becomes the usual Dirac $\delta$ function as $\sigma_x\to 0$. We have written $N(t,\omega)$ for the number of tips at time $t$ to emphasize that this number may be different for different replicas. Similarly, the stochastic distribution of tips per unit volume in the physical space at time $t$ is
\begin{equation}
\tilde{Q}_N^*(t,\mathbf{x}, \omega)=
\sum_{i=1}^{N(t,\omega)}\delta_{\sigma_x}(\mathbf{x}-\mathbf{X}^i(t,\omega))\!.
\label{eq8}
\end{equation}
As a  consequence
\begin{equation}
\delta_{\sigma_x}(\mathbf{x}-\mathbf{X}(t,\omega))=\int_0^t\sum_{i=1}^{N(s,\omega)}\delta_{\sigma_x}(\mathbf{x}-\mathbf{X}^i(s,\omega)) ds, \label{eq9}
\end{equation}
represents the concentration of all vessels per unit volume in the physical space, at time $t$, i.e., the vessel network.

In \cite{bon14}, we assumed that the number of tips $N(t,\omega)$ could be sufficiently large for a scaled version of (\ref{eq7}) to converge to a density of tips in phase space. Our numerical simulations show that anastomosis keeps $N(t,\omega)$ moderate and therefore we need to follow a different path to define a tip density. Considering $\mathcal{N}$ replicas of the angiogenic process, we define the empirical distribution of tips per unit volume in the $(\mathbf{x},\mathbf{v})$ phase space,
\begin{eqnarray}
p_{\mathcal{N}}\!(t,\mathbf{x},\mathbf{v})\!&=&\!\frac{1}{\mathcal{N}}\sum_{\omega=1}^\mathcal{N}\sum_{i=1}^{N(t,\omega)}\delta_{\sigma_x}(\mathbf{x}-\mathbf{X}^i(t,\omega))
%\nonumber\\&\times& 
\delta_{\sigma_v}(\mathbf{v}-\mathbf{v}^i(t,\omega))\nonumber\\
\!&=&\!\frac{1}{\mathcal{N}}\sum_{\omega=1}^\mathcal{N} Q_N^*(t,\mathbf{x},\mathbf{v},\omega), \label{eq10}
\end{eqnarray}
and, correspondingly, the empirical distribution of tips per unit volume in the physical space and the vessel tip flux are, 
\begin{eqnarray}
\tilde{p}_{\mathcal N}(t,\mathbf{x})\!\!&=&\!\frac{1}{\mathcal{N}}\sum_{\omega=1}^\mathcal{N}\sum_{i=1}^{N(t,\omega)}\delta_{\sigma_x}(\mathbf{x}-\mathbf{X}^i(t,\omega)), \label{eq11}\\
\mathbf{j}_{\mathcal N}(t,\mathbf{x})\!\!&=&\!\frac{1}{\mathcal{N}}\!\sum_{\omega=1}^\mathcal{N}\!\sum_{i=1}^{N(t,\omega)}\!\!\mathbf{v}^i(t,\omega)\delta_{\sigma_x}(\mathbf{x}-\mathbf{X}^i(t,\omega)),\label{eq12}
\end{eqnarray}
respectively. Thus what we propose is a new deterministic description based upon using the classical law of large numbers on the arithmetic mean over a large number $\mathcal{N}$ of independent replicas.

Now assuming that $\sigma_x$ and $\sigma_v$ go to zero as the number of replicas goes to infinity, %we assume that 
the following limit exists 
\begin{eqnarray}
\!\!\!\!p(t,\mathbf{x},\mathbf{v})\!\!\!&=&\!\!\!\lim_{\mathcal{N}\to\infty}
p_{\mathcal{N}}(t,\mathbf{x},\mathbf{v}) %\nonumber \\\!\!&=&\!\!
=\!\left\langle
\sum_{i=1}^{N(t,\cdot)}\!\!\delta_{\sigma_x}(\mathbf{x}-\mathbf{X}^i(t,\cdot))\delta_{\sigma_v}(\mathbf{v}-\mathbf{v}^i(t,\cdot))\!\!\right\rangle\!\!.
\label{eq13}
\end{eqnarray}
We assume that $p(t,\mathbf{x},\mathbf{v})$ is the deterministic distribution of tips per unit phase space volume. The proof of this statement is not trivial (and out of the scope of this paper), as the usual assumptions on the kernel density estimation (see page 489 of \cite{roussas}) may not apply. Similarly the deterministic distribution of tips per unit volume in physical space should exist as the following limit
 \begin{eqnarray}
\tilde{p}(t,\mathbf{x})&=&\lim_{\mathcal{N}\to\infty}
\tilde{p}_{\mathcal N}(t,\mathbf{x}) =%\nonumber\\&=&
\!\left\langle
\sum_{i=1}^{N(t,\cdot)}\delta_{\sigma_x}(\mathbf{x}-\mathbf{X}^i(t,\cdot))\right\rangle\!.
\label{eq14}
\end{eqnarray}
Finally, we may obtain the deterministic version of the vessel tip flux as 
\begin{eqnarray}
\mathbf{j}(t,\mathbf{x})\!\!&=&\!\!\lim_{\mathcal{N}\to\infty}\mathbf{j}_{\mathcal N}(t,\mathbf{x})%\nonumber\\\!&=&\!\!
=\!\left\langle \sum_{i=1}^{N(t,\cdot)}\mathbf{v}^i(t,\cdot)\,\delta_{\sigma_x}(\mathbf{x}-\mathbf{X}^i(t,\cdot))\right\rangle\!.
\label{eq15}
\end{eqnarray}

\begin{figure}[ht]
\begin{center}
\includegraphics[width=11cm]{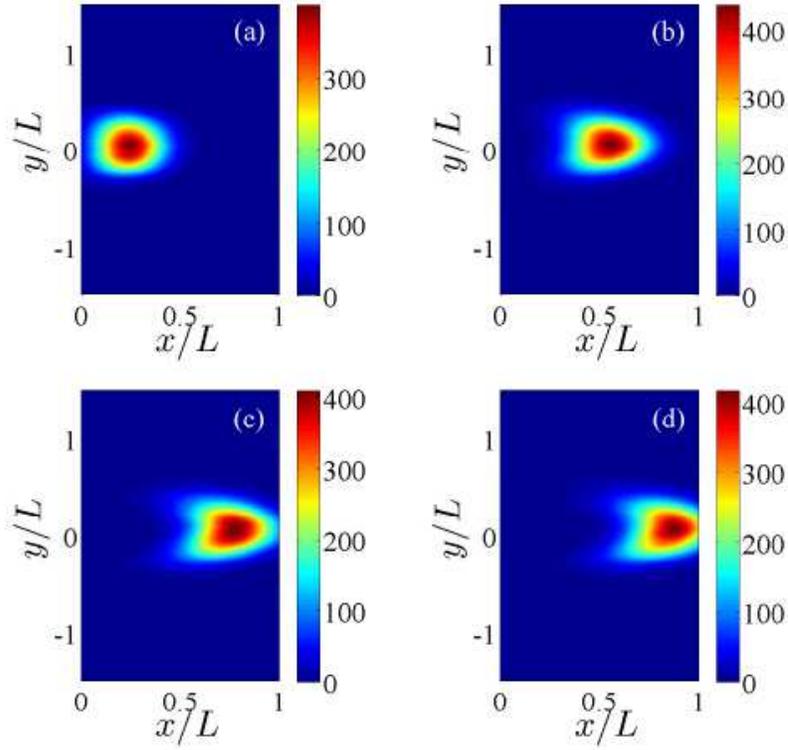} \qquad
\end{center}
\caption{(Color online) Density plot of the marginal tip density $\tilde{p}(t,x,y)$ calculated from (\ref{eq11}) with $\mathcal{N}=400$ replicas for the same times as in Figure \ref{fig1} showing how tips are created at $x=0$ and march towards the tumor at $x=L$. At these times, the number of active tips are (a) 56, (b) 69, (c) 72, and (d) 66.
\label{fig2}}
\end{figure}

\begin{figure}[ht]
\begin{center}
\includegraphics[width=11cm]{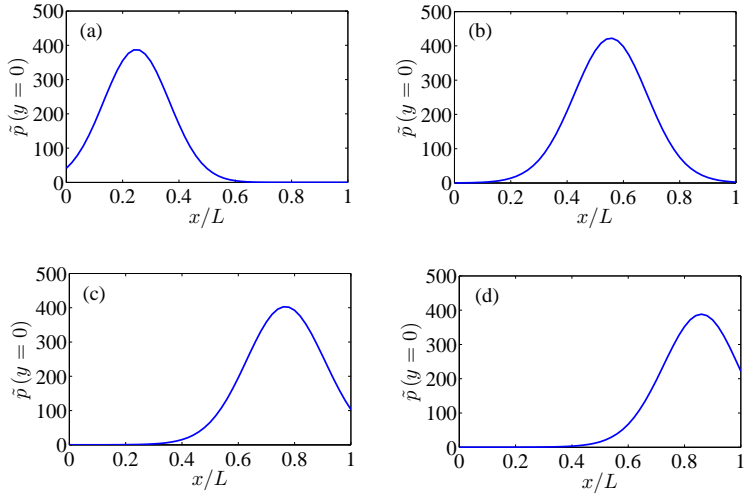} \qquad
\end{center}
\caption{(Color online) Marginal tip density at the $x$ axis, $\tilde{p}(t,x,y=0)$, calculated by averages over 400 replicas of the stochastic description for the same times as in Figure \ref{fig1}. A pulse of the marginal tip density is created at the primary vessel and marches towards the tumor. 
\label{fig4}}
\end{figure}

Figure \ref{fig2} shows the marginal tip density $\tilde{p}(t,x,y)\approx\tilde{p}_{\mathcal N}(t,x,y)$ calculated from (\ref{eq11}) with $\mathcal{N}=400$ replicas at the same times represented in Figure \ref{fig1}. Ensemble averages over a larger number of replicas are the same. We observe that the tips proliferate after a few hours and reach a high number by branching onto the free space ahead of them. Influenced by chemotaxis, the marginal tip density thickens about the $x$ axis and it forms a lump that advances toward the tumor.  Behind the lump, the density drops to a low value. While the network of vessels is formed and is quite dense in the wake of the tips (as shown by Figure \ref{fig1}), the active tips diminish by anastomosis there and they are numerous only at the leading part of the lump where free space is available. This is made clearer by plotting the marginal tip density at the $x$ axis as in Figure \ref{fig4} for the same times as in Figures \ref{fig1} and \ref{fig2}. We observe that the marginal tip density at a point decreases to very small values after the lump of tips passage. This is another indication that the definition of  marginal tip density based on ensemble average provides a better deterministic description of angiogenesis than the density based on using the law of large numbers on a single replica: there are no tips or very few ones in large regions of the physical space where the law of large numbers is inapplicable. 

\section{Deterministic description}
\label{sec:deterministic}
In Ref.~\cite{bon14}, the law of large numbers (the\,Ê{\em propagation of molecular chaos} assumption) was used to derive an integrodifferential equation of Fokker-Planck type for the vessel tip density $p(t,\mathbf{x},\mathbf{v})$ coupled to a reaction-diffusion equation for the TAF concentration. In this Section, we give a different derivation of the {\em Fokker-Planck type}  deterministic equation of Ref.~\cite{bon14}, with the new interpretation  of the tip density given by Equation (\ref{eq13}). By following a similar approach as in the Appendix of our previous paper \cite{bon14}, we may obtain the weak formulation of the stochastic evolution of $Q_N^*(t,\mathbf{x},\mathbf{v},\omega)$, defined in (\ref{eq7}), which is the same as Equation (A13) there:
%\begin{widetext}
\begin{eqnarray}
&&\int\! g(\mathbf{x},\mathbf{v})\,
Q_N^*(t,\mathbf{x},\mathbf{v})d\mathbf{x} d\mathbf{v} = \!\int\!
g(\mathbf{x},\mathbf{v})\,
Q_N^*(0,\mathbf{x},\mathbf{v})d\mathbf{x} d\mathbf{v}\nonumber\\
&&+\int_0^t\!\int\!\mathbf{v}\cdot\nabla_xg(\mathbf{x},\mathbf{v})
Q_N^*(s,\mathbf{x},\mathbf{v})d\mathbf{x} d\mathbf{v}\, ds \nonumber\\
&&\, +\!\! \int_0^t\!\!\!\int\!
[\mathbf{F}(C(s,\mathbf{x}))-k\mathbf{v}]\!\cdot\!\nabla_v
g(\mathbf{x},\mathbf{v}) Q_N^*(s,\mathbf{x},\mathbf{v})d\mathbf{x}
d\mathbf{v}\, ds \nonumber\\
&&+ \int_0^t\!\int\! \frac{\sigma^2}{2}\Delta_v
g(\mathbf{x},\mathbf{v})
Q_N^*(s,\mathbf{x},\mathbf{v})d\mathbf{x} d\mathbf{v} ds\nonumber\\
&&\, + \int_0^t \!
\!\int\!\alpha(C(s,\mathbf{x}))\delta_{\sigma_v}\!(\mathbf{v}-\mathbf{v}_0)Q_N^*(s,\mathbf{x},\mathbf{v})d\mathbf{x}
d\mathbf{v}\, ds \nonumber\\
&&- \gamma\! \int_0^t\!\! \int\!
\delta_{\sigma_x}\!(\mathbf{x}-\mathbf{X}(s))
g(\mathbf{x},\mathbf{v}) Q_N^*(s,\mathbf{x},\mathbf{v})d\mathbf{x}
d\mathbf{v}\, ds + \tilde{M}_N\!(t).\label{eq16}
\end{eqnarray}
%\end{widetext}
Here $g(\mathbf{x},\mathbf{v})$ is a smooth test function, $\gamma>0$ is a parameter characterizing the anastomosis, and $\tilde{M}_N(t)$ is a zero mean martingale, which collects the source of randomness  of the system \cite{bon14}. According to (\ref{eq13}), the law of large numbers applied to the arithmetic mean may produce $\langle Q_N^*(t,\mathbf{x},\mathbf{v})\rangle\sim p(t,\mathbf{x},\mathbf{v})$ if the limit of the density for infinitely many replicas exist. Furthermore, we use the approximation (which becomes exact if the law of large numbers is applicable):
\begin{eqnarray}
&&\!\left\langle\sum_{i=1}^{N(t,\cdot)}\!\!\mathbf{F}(C(t,\mathbf{X}^i(t,\cdot)))\delta_{\sigma_x}(\mathbf{x}-\mathbf{X}^i(t,\cdot))\delta_{\sigma_v}(\mathbf{v}-\mathbf{v}^i(t,\cdot))\!\right\rangle\!\!\nonumber\\
&&\approx\!\mathbf{F}(\langle C(t,\mathbf{x})\rangle) \!\left\langle\sum_{i=1}^{N(t,\cdot)}\!\!\delta_{\sigma_x}(\mathbf{x}-\mathbf{X}^i(t,\cdot))\delta_{\sigma_v}(\mathbf{v}-\mathbf{v}^i(t,\cdot))\!\right\rangle%\nonumber\\&&
=\mathbf{F}(\langle C(t,\mathbf{x})\rangle)\, p(t,\mathbf{x},\mathbf{v}),  \label{eq17}
\end{eqnarray}
Hence, on the basis of the above convergence assumptions and $\langle\tilde{M}_N(t)\rangle=0$, we may expect that the ensemble average of the stochastic equation (\ref{eq16}) tends in its strong form to the same equation of Fokker-Planck type as in \cite{bon14}:
\begin{eqnarray}
\frac{\partial}{\partial t} p(t,\mathbf{x},\mathbf{v})\!&=&\!
\frac{\alpha_1C(t,\mathbf{x})}{C_R+C(t,\mathbf{x})}\,
 p(t,\mathbf{x},\mathbf{v})\delta_{\sigma_v}(\mathbf{v}-\mathbf{v}_0)
 - \gamma p(t,\mathbf{x},\mathbf{v}) \int_0^t \tilde{p}(s,\mathbf{x})\, ds  - \mathbf{v}\cdot \nabla_x   p(t,\mathbf{x},\mathbf{v}) \nonumber\\ 
 &+& k \nabla_v \cdot [\mathbf{v} p(t,\mathbf{x},\mathbf{v})] - d_1\nabla_v \cdot
\left[\frac{\nabla_x C(t,\mathbf{x})}{[1+\gamma_1C(t,\mathbf{x})]^q}\, p(t,\mathbf{x},\mathbf{v})
\right]\! + \frac{\sigma^2}{2} \Delta_{v} p(t,\mathbf{x},\mathbf{v}).
\label{eq18}
\end{eqnarray}
Here the marginal vessel tip density,
\begin{equation}
\tilde{p}(t,\mathbf{x}) = \int p(t,\mathbf{x},\mathbf{v}')\, d\mathbf{v}',\label{eq19}
\end{equation} 
is the marginal density of $p(t,\mathbf{x},\mathbf{v})$. We couple  Equation (\ref{eq18}) for  $p(t,\mathbf{x},\mathbf{v})$ with a deterministic reaction-diffusion equation for the TAF concentration,
\begin{eqnarray} 
\frac{\partial}{\partial t}C(t,\mathbf{x})=d_2 \Delta_x C(t,\mathbf{x})- \eta\, C(t,\mathbf{x})\!\left| \mathbf{j}(t,\mathbf{x})\right|\!,\label{eq20}
\end{eqnarray}
where $\mathbf{j}(t,\mathbf{x})$ is the ensemble-averaged current density (flux) vector at any point $\mathbf{x}$ and any time $t\geq 0$,
\begin{equation}
\mathbf{j}(t,\mathbf{x})= \int \mathbf{v}' p(t,\mathbf{x},\mathbf{v}')\, d \mathbf{v'}. \label{eq21}
\end{equation}
On the right hand side of (\ref{eq18}), the first (birth) term is the ensemble average of (\ref{eq6}) per unit phase space volume. The second (death) term is proportional to the occupation time density $\int_0^t\tilde{p}(s,\mathbf{x}) ds$. The fraction of time a small volume $d\mathbf{x}$ about $\mathbf{x}$ is occupied by tips, no matter their velocity and realization of the stochastic process, is the ensemble average of (\ref{eq9}):
\begin{equation}
\int_0^t\left\langle\sum_{i=1}^{N(s,\cdot)}\delta_{\sigma_x}(\mathbf{x}-\mathbf{X}^i(s,\cdot))\right\rangle ds=\int_0^tÊ\tilde{p}(s,\mathbf{x}) ds, \label{eq22}
\end{equation}
per unit volume \cite{karlin2}. In angiogenesis, tips occupy the volume $d\mathbf{x}$ about $\mathbf{x}$ either the first time they reach it, or by branching or during anastomosis. Thus the average occupation time density during anastomosis should be a fraction of (\ref{eq22}). While anastomosis of one tip depends on the past history of the considered replica (i.e., realization) of the stochastic process, the ensemble average involved in the definition of the vessel tip density takes into account all possible replicas that are, by definition, independent. Then we expect the death term in (\ref{eq18}) to be proportional to the occupation time density (\ref{eq22}), which is just the second term on the right hand side of (\ref{eq18}). For appropriate initial and boundary data, it is possible to prove that (\ref{eq18}) and (\ref{eq20}) have a unique smooth solution \cite{ana}.

\paragraph{Boundary and initial conditions}
%\label{sec:bc}
We solve the system of equations (\ref{eq18}) and (\ref{eq20}) in a two dimensional strip geometry using the initial and boundary conditions introduced in \cite{bon14}. The strip is $\Omega=[0,L]\times\mathbb R \subset \mathbb R^2$, its left boundary $\Omega_0=(0,y)$, $y\in\mathbb R$ is the primary vessel issuing new vessels, and $\Omega_L=(L,y)$, $y\in\mathbb R$, includes the tumor which is a source of  the TAF $C.$ Let $c_1(y)$ be the TAF flux emitted by the tumor at $x=L$. The boundary conditions for the TAF are 
\begin{eqnarray}
&&\frac{\partial}{\partial x} C(t,0,y)=0, \quad
\frac{\partial}{\partial x} C(t,L,y)=\frac{c_1(y)}{d_2}=\frac{a}{d_2} e^{-y^2/b^2}\label{eq23}
\end{eqnarray}
($b$ is half the tumor width), and $C\to 0$ as $|y|\to\infty$. We do not intend to follow the process of angiogenesis beyond the time that vessels tip have arrived at the tumor and therefore we do not give the latter a finite length. We use a Gaussian as the initial condition for the TAF 
\begin{eqnarray}
C(0,x,y)=1.1\, C_R e^{-[(x-L)^2/c^2+y^2/b^2]},\label{eq24}
\end{eqnarray} 
for appropriate $b$ and $c$. 

The boundary conditions for the tip density are
\begin{eqnarray}
&& p^+(t,0,y,v,w)
=\frac{e^{-\frac{k|\mathbf{v}-\mathbf{v}_0|^2}{\sigma^2}}}{\int_0^{\infty}\!\int_{-\infty}^{\infty}
 v' e^{-\frac{k|\mathbf{v}'-\mathbf{v}_0|^2}{\sigma^2}}dv'\,dw'} \nonumber \\
&& \times\!  \left[j_0(t,y)\! -\! \int_{-\infty}^0\!\int_{-\infty}^{\infty}\!\!
v' p^-(t,0,y,v',w')d v' dw'\right]\!\!, \label{eq25} \\
&&p^-(t,L,y,v,w)=\frac{e^{-\frac{k|\mathbf{v}-\mathbf{v}_0|^2}{\sigma^2}}}{\int_{-\infty}^0\!\int_{-\infty}^{\infty}
e^{-\frac{k|\mathbf{v}'-\mathbf{v}_0|^2}{\sigma^2}}dv'\,dw'} \nonumber\\
&& \times  \!\left[\tilde{p}(t,L,y)\! -\!\int_0^{\infty}\!\!\int_{-\infty}^{\infty}\! p^+(t,L,y,v',w')dv' dw'\!\right]\!\!, \label{eq26}\\
&&p(t,\mathbf{x},\mathbf{v})\to 0 \mbox{ as } |\mathbf{v}|\to \infty,\label{eq27}
\end{eqnarray}
where $p^+=p$ for $v>0$ and $p^-=p$ for $v<0$, $\mathbf{v}=(v,w)$. The tip flux density at $x=0$ is \cite{bon14}
\begin{equation}
j_0(t,y)=\frac{v_0\, L}{\sqrt{v_0^2+w_0^2}}\alpha(C(t,0,y))\, p(t,0,y,v_0,w_0), \label{eq28}
\end{equation}
for the vector velocity $\mathbf{v}_0=(v_0,w_0)$. The boundary condition \eqref{eq25} implies that the vessel tip flux at the primary vessel is \eqref{eq28}, built out from the tip branching probability. The boundary condition \eqref{eq26} is compatible with the instantaneous value of the tip marginal density at $x=L$. This condition considers that all tips arriving at $x=L$ have reached the tumor. Thus the total number of active tips produced by the deterministic equations should be smaller than the total number of tips provided by the stochastic process once the first tips arrive at $x=L$.

The initial condition for the tip density is 
\begin{eqnarray}
p(0,x,y,v,w)&=& \frac{e^{-x^2/l_x^2}}{\pi^{3/2}l_x\sigma_v^2}\,e^{-|\mathbf{v}-\mathbf{v}_0|^2/\sigma_v^2}%\nonumber\\&\times&
\sum_{i=1}^{N_0}\frac{1}{\sqrt{\pi}l_y} (e^{-|y-y_i|^2/l_y^2}+e^{-|y+y_i|^2/l_y^2}).\label{eq29}
\end{eqnarray} 
As $l_x$ and $l_y$ tend to zero, (\ref{eq29}) becomes 
\begin{eqnarray}
p(0,x,y,\mathbf{v})\!=\! \delta_{\sigma_v}(\mathbf{v}-\mathbf{v}_0)\delta(x)\sum_{i=1}^{N_0}[\delta(y-y_i)+\delta(y+y_i)].\label{eq30}
\end{eqnarray} 
This initial condition corresponds to the following initial condition for the stochastic process: There are $N_0$ equally spaced initial tips at $x=0$, with vertical positions $\pm y_i$ equally spaced on the interval $[-L_y,L_y]$, whose initial velocities are normally distributed about $\mathbf{v}_0$ with standard deviation $\sigma_v$. 

\section{Numerical results and determination of the anastomosis coefficient}
\label{sec:numerical}
We use the parameter values indicated in Table \ref{table} that have been extracted from experiments as explained in Ref. \cite{bon14}. The anastomosis coefficient $\gamma$ was given an arbitrary value in \cite{bon14}, whereas we estimate it here so as to get a good agreement between simulations of the stochastic equations and solutions of the deterministic equations. We nondimensionalize the governing equations of our model, (\ref{eq18}) and (\ref{eq20}), according to the units in Table \ref{table1}, thereby obtaining 

\begin{table}[ht]
\begin{center}\begin{tabular}{cccccccc}
 \hline
$\frac{1}{k}$& $\tilde{v}_0$ & $\sigma^2$&$\alpha_1$&$d_1C_R$ &$C_R$ &$\eta$ &$\gamma$\\
hr & $\frac{\mbox{$\mu$m}}{\mbox{hr}}$ & $10^{-21}\frac{\mbox{m}^2}{\mbox{s}^3}$& $10^{-20}\frac{\mbox{m}^2}{\mbox{s}^3}$ & $\frac{\mbox{$\mu$m$^2$}}{\mbox{hr}^2}$  & mol/m$^2$  & $\mu$m& $10^{-17}\frac{\mbox{m$^2$}}{\mbox{s$^2$}}$\\ 
8.5& 40 &4.035 & 1.538& 2400 & $10^{-16}$ & 4& 1.79\\
 \hline
\end{tabular}
\end{center}
\caption{Parameters used to solve the model equations. }
\label{table}
\end{table}

\begin{table}[ht]
\begin{center}\begin{tabular}{ccccccc}
 \hline
$\mathbf{x}$& $\mathbf{v}$ & $t$ &$C$& $p$ &$\tilde{p}$&$\mathbf{j}$\\
$L$ & $\tilde{v}_0$ & $\frac{L}{\tilde{v}_0}$ & $C_R$ & $\frac{1}{\tilde{v}_0^2L^2}$& $\frac{1}{L^2}$& $\frac{\tilde{v}_0}{L^2}$\\
mm&$\mu$m/hr & hr& mol/m$^2$&$10^{21}\frac{\mbox{s$^2$}}{\mbox{m$^4$}}$ & $10^{5}$m$^{-2}$&m$^{-1}$s$^{-1}$\\
$2$& 40 & 50 & $10^{-16}$ & 2.025 & 2.5 & 0.0028 \\ 
\hline
\end{tabular}
\end{center}
\caption{Units for nondimensionalizing the model equations. }
\label{table1}
\end{table}

\begin{eqnarray} 
\frac{\partial p}{\partial t} &=& \frac{A\, C}{1+C}p\, \delta_v(\mathbf{v}-\mathbf{v}_0)
- \Gamma p\! \int_0^t \tilde{p}(s,\mathbf{x})\, ds  %\nonumber\\&-& 
- \mathbf{v}\cdot \nabla_x   p  \nonumber\\&-&
\nabla_v \cdot\left[\!\left(\frac{\delta\, \nabla_x C}{(1+\Gamma_1C)^q}-\beta\mathbf{v}\right)\! p\right]  %\nonumber \\ &+&  
+\frac{\beta}{2}\, \Delta_{v} p,\label{eq31}\\
\frac{\partial C}{\partial t} &=& \kappa\, \Delta_x C - \chi\,
C\, |\mathbf{j}| \label{eq32}.
\end{eqnarray}
The dimensionless parameters appearing in these equations are defined in Table \ref{table2}. We have used $\sigma_v^2=\sigma^2\epsilon^2/k$ in (\ref{eq4}) as the variance in the Gaussian function $\delta_{\sigma_v}(\mathbf{v})$, and obtained the nondimensional function,
\begin{eqnarray}
\delta_{v}(\mathbf{v})= \frac{1}{\pi\epsilon^2}e^{-|\mathbf{v}|^2/\epsilon^2}, \label{eq33}
\end{eqnarray}
that appears in (\ref{eq31}). We have used $\epsilon=0.08$ and $q = 1$ in our numerical simulations.

\begin{table}[ht]
\begin{center}\begin{tabular}{ccccccc}
 \hline
$\delta$ & $\beta$ &$A$& $\Gamma$& $\Gamma_1$ &$\kappa$&$\chi$\\
 $\frac{d_1C_R}{\tilde{v}_0^2}$ & $\frac{kL}{\tilde{v}_0}$ & $\frac{\alpha_1L}{\tilde{v}_0^3}$ & $\frac{\gamma}{\tilde{v}_0^2}$&$\gamma_1C_R$& $\frac{d_2}{\tilde{v}_0 L}$& $\frac{\eta}{L}$\\
1.5 & 5.88 & $22.42$ & 0.145 & 1& $0.0045$ & 0.002 \\
 \hline
\end{tabular}
\end{center}
\caption{Dimensionless parameters. } 
\label{table2}
\end{table}

The nondimensional boundary conditions for $C$,
\begin{eqnarray}
\frac{\partial C}{\partial x}(t,0,y)=0, \quad\frac{\partial C}{\partial x}(t,1,y)=f(y), \, \lim_{y\to\pm\infty}C= 0, \label{eq34}
\end{eqnarray}
where $f(y)=L\, c_1(Ly)/(C_Rd_2)$ is a nondimensional flux, follow from (\ref{eq23}). We have used $c_1(y)=a\, e^{-y^2/b^2}$, with $a=5.5\times 10^{-27}$ mol/(m\, s), $d_2=10^{-13}$ m$^2$/s, and $b=0.6$ mm ($b$ is about half the assumed tumor size). The initial condition for the TAF (\ref{eq24}) yields
\begin{eqnarray}
C(0,x,y)=1.1\, e^{-[(x-1)^2L^2/c^2+y^2L^2/b^2]},\label{eq35}
\end{eqnarray} 
with $b/L=0.3$, $c/L=1.5$, whereas the nondimensional initial vessel density is 
\begin{eqnarray}
p(0,x,y,v,w)&=& \frac{L e^{-x^2 L^2/l_x^2}}{\pi^{3/2}l_x\epsilon^2}\,e^{-|\mathbf{v}-\mathbf{v}_0|^2/\sigma_{v}^2}%\nonumber\\&\times&
\sum_{i=1}^{N_0}\frac{L}{\sqrt{\pi}l_y} (e^{-|y-y_i|^2 L^2/l_y^2}+e^{-|y+y_i|^2 L^2/l_y^2}),
\label{eq36}
\end{eqnarray} 
with $l_x/L=0.06$ and $l_y/L=0.08$, that corresponds to $N_0=20$ initial vessel tips. In nondimensional form, the boundary conditions (\ref{eq25})-(\ref{eq26}) for $p$ are
\begin{eqnarray}
&& p^+(t,0,y,v,w)
=\frac{e^{-|\mathbf{v}-\mathbf{v}_0|^2}}{\int_0^{\infty}\!\int_{-\infty}^{\infty}
 v' e^{-|\mathbf{v}'-\mathbf{v}_0|^2} dv'\,dw'} \nonumber \\
 && \times\! \left[j_0(t,y)\! -\! \int_{-\infty}^0\!\int_{-\infty}^{\infty}\!\!
v' p^-(t,0,y,v',w')d v' dw'\right]\!  \label{eq37}
 \end{eqnarray}
for $x=0$ and $v>0,$
\begin{eqnarray}
&&p^-(t,1,y,v,w)=\frac{e^{-|\mathbf{v}-\mathbf{v}_0|^2}}{\int_{-\infty}^0\!\int_{-\infty}^{\infty}
e^{-|\mathbf{v}'-\mathbf{v}_0|^2}dv'\,dw'} \nonumber\\
&& \times  \!\left[\tilde{p}(t,1,y)\! -\!
\int_0^{\infty}\!\!\int_{-\infty}^{\infty}\! p^+(t,1,y,v',w')dv' dw'\right]\!
\label{eq38}
\end{eqnarray}
for $x=1$ and $v<0$. Eq.\! (\ref{eq28}) produces the nondimensional flux $j_0$:
\begin{eqnarray}
j_0(t,y)=A\, v_0\frac{C}{1+C}\, p(t,0,y,v_0,w_0)   \label{eq39}
\end{eqnarray}
($\sqrt{v_0^2+w_0^2}=|\mathbf{v}_0|^2=1$ in nondimensional units).

\begin{figure}[ht]
\begin{center}
\includegraphics[width=11cm]{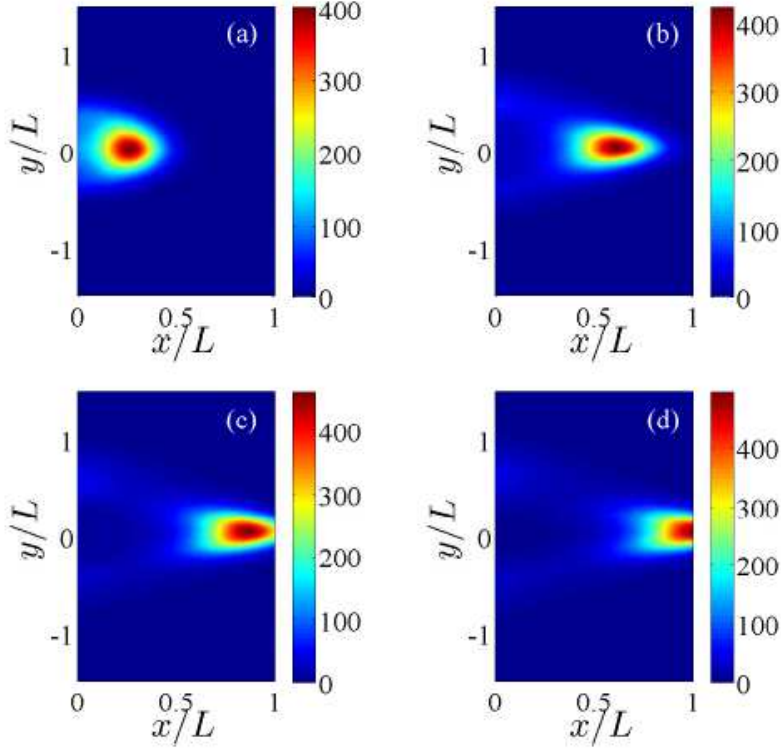} \qquad
\end{center}
\vskip -1cm
\caption{(Color online) Density plots of the marginal tip density calculated from the deterministic description for the same times as in Figure \ref{fig2}. At these times, the number of active tips are (a) 57, (b) 68, (c) 71, and (d) 61.
\label{fig3}}
\end{figure}

As in \cite{bon14}, we have solved (\ref{eq31})-(\ref{eq39}) by an explicit finite-difference scheme, using upwind differences for positive $v$ and $w$ and downwind differences for negative $v$ and $w$. (\ref{eq37}) and (\ref{eq38}) give the needed boundary value of $p^\pm$ at one time step in terms of the value of $p^\mp$, which is known at the precedent time step. The integrals are approximated by the composite Simpson rule. In \cite{bon14}, we showed the consistency of the deterministic model by depicting TAF concentration, marginal tip density and overall network density at different times. In this paper, we have chosen the marginal tip density to compare deterministic and stochastic descriptions. The evolution of the marginal tip density has been calculated by averaging over 400 realizations of the stochastic description in Figure \ref{fig2} and by numerically solving the deterministic problem (\ref{eq31})-(\ref{eq39}) in Figure \ref{fig3}. These figures show that both descriptions agree quite well for the anastomosis coefficient we have selected (see below).

\begin{figure}[ht]
\begin{center}
\includegraphics[width=11cm]{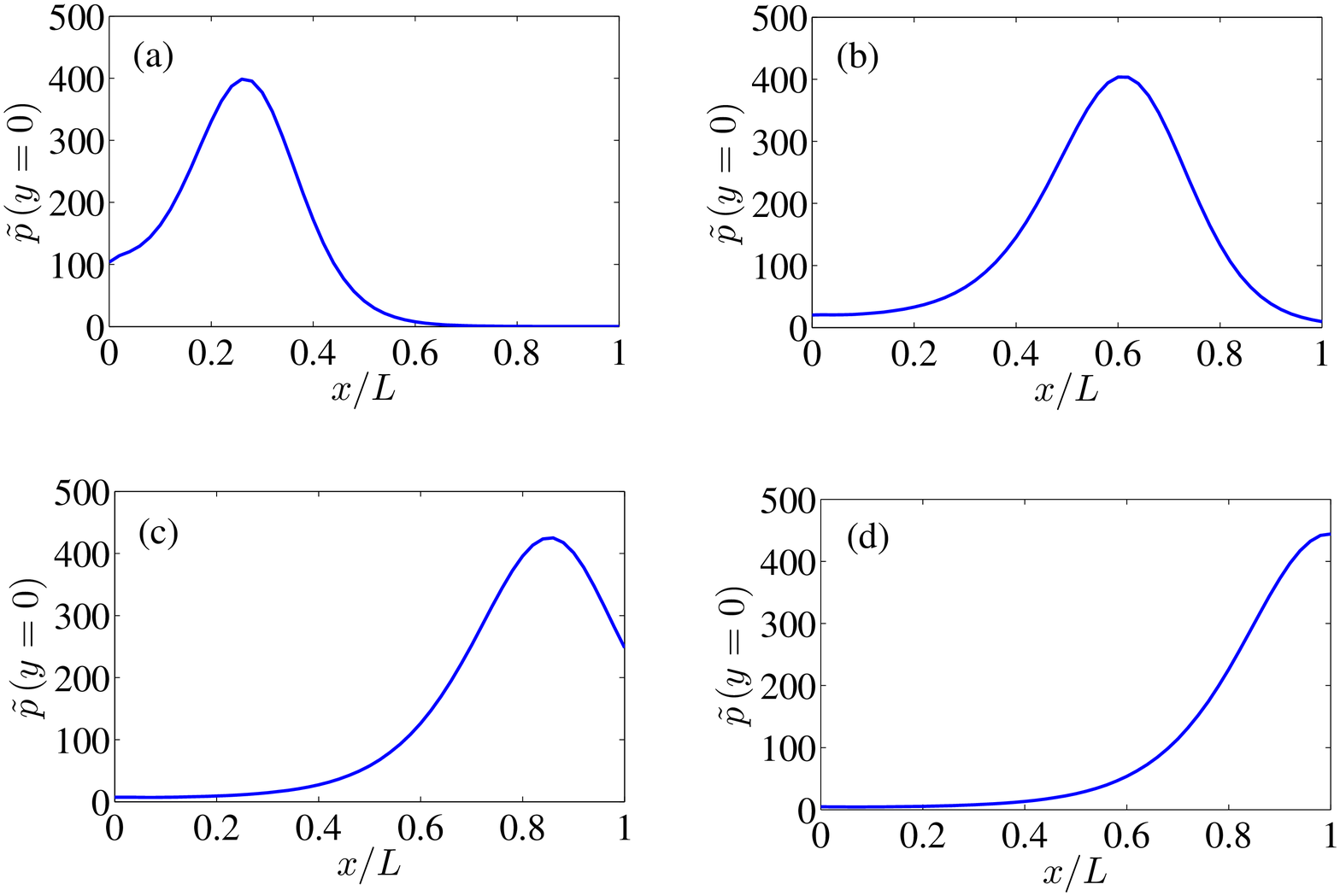}\qquad
\end{center}
\vskip -1cm
\caption{(Color online) Same as Figure \ref{fig4} but now the marginal tip density at the $x$ axis, $\tilde{p}(t,x,y=0)$, is calculated from the deterministic description.
\label{fig5}}
\end{figure}

\begin{figure}[ht]
\begin{center}
\includegraphics[width=9cm]{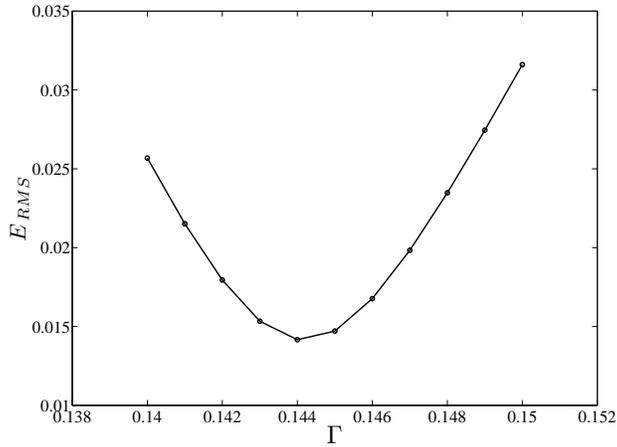} \qquad
\end{center}
\vskip -1cm
\caption{(Color online) RMS error between ensemble average number of tips (over 400 replicas) and number of tips calculated from the deterministic description based on integrodifferential equations. $t_o$ and $t_f$ are 8 and 30 hours, respectively.
\label{fig8}}
\end{figure}

If we plot the marginal tip density at the $x$ axis, $\tilde{p}(t,x,y=0)$, ensemble averages of the stochastic description (Figure \ref{fig4}) and numerical solution of the deterministic description (Figure \ref{fig5}) show that the vessel tips form a growing pulse that moves to the tumor by chemotaxis. The total number of tips, $N(t)$, is counted as the integer part of the integral $\int\tilde{p}(t,\mathbf{x})\, d\mathbf{x}$. There are small discrepancies between stochastic and deterministic descriptions that are more appreciable as the tips arrive at the tumor at $x=L$. The behavior of the angiogenic vessel network depends very much on the values of the dimensionless parameters in Table \ref{table2}. We have selected the anastomosis rate, $\gamma$, in such a way that the number of tips at each time, $N(t)$, evolves similarly for the deterministic description based on integrodifferential equations and for the ensemble averages of $N(t,\omega)$ calculated from the stochastic simulations. In Figure \ref{fig8}, we have depicted the root mean square (RMS) error between $\int\tilde{p}(t,\mathbf{x})\, d\mathbf{x}$ (calculated for different values of the anastomosis coefficient) and $\langle N(t,\cdot)\rangle$: 
\begin{eqnarray}
E_{\rm RMS}=\sqrt{\frac{\int_{t_o}^{t_f} |N(t;\Gamma)- \langle N(t,\cdot)\rangle|^2dt}{\int_{t_o}^{t_f}|\langle N(t,\cdot)\rangle|^2 dt} },\quad N(t;\Gamma)=\int\tilde{p}(t,\mathbf{x};\Gamma)\, d\mathbf{x}.   \label{eq40}
\end{eqnarray}

We use a dimensionless anastomosis coefficient $\Gamma=0.145$, that corresponds to $\gamma = 1.79 \times 10^{-17}$m$^2$/s$^2$, and a 1.5\% error. Selecting values of $\Gamma$ close to 0.145 yields quite similar results.

\begin{figure}[ht]
\begin{center}
\includegraphics[width=9cm]{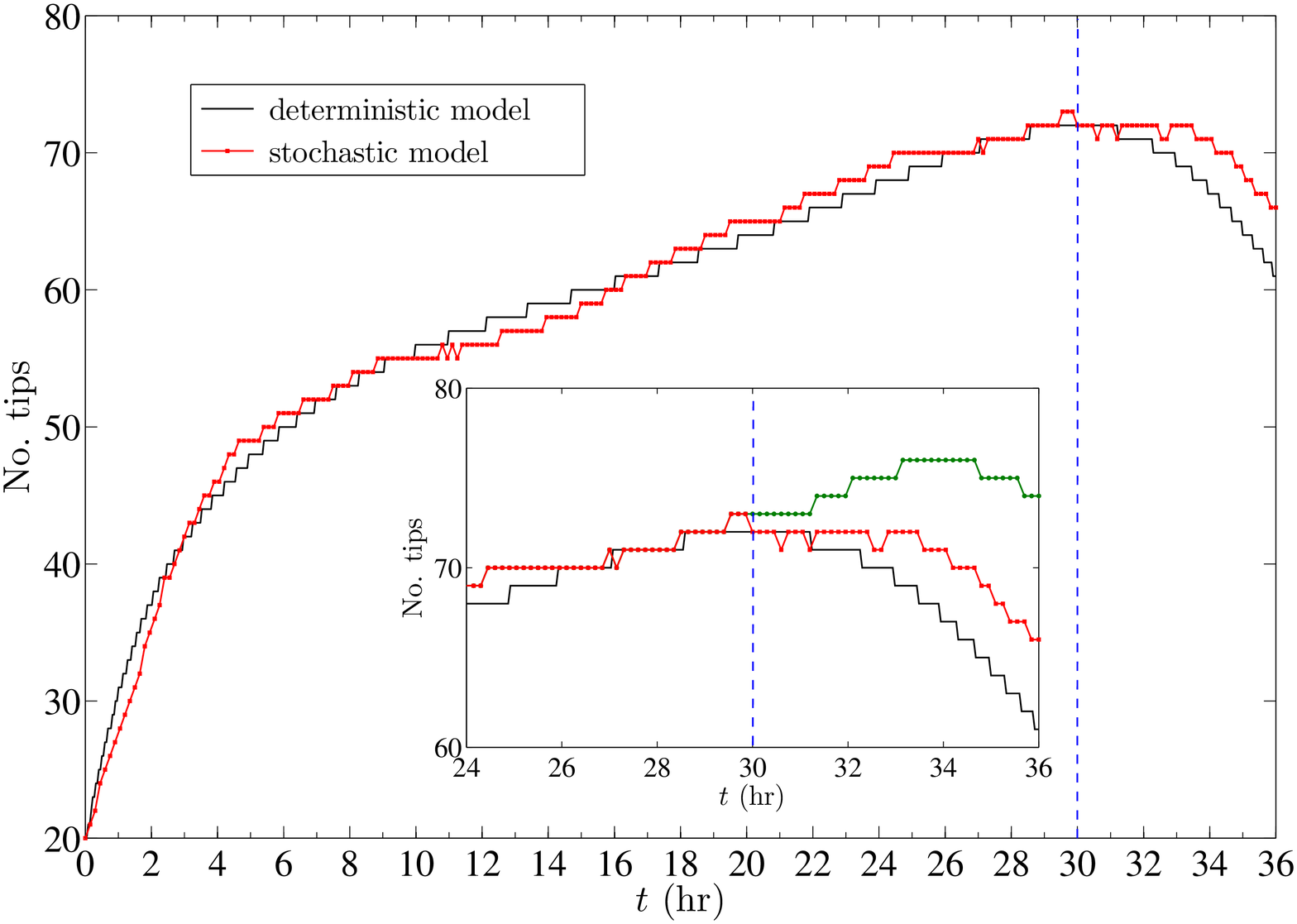} \qquad
\end{center}
\vskip -1cm
\caption{(Color online) Evolution of the number of tips as calculated from the deterministic description based on integrodifferential equations (solid black line), and from ensemble averages over 400 realizations of the stochastic description (solid red line). Inset: evolution at later times shows that directly counting tips and ensemble averaging (solid green line) gives a larger number than integrating the marginal tip density over space. 
\label{fig6}}
\end{figure}

In Figure \ref{fig6}, we have depicted the total number of tips as the integer parts:
\begin{eqnarray}
&&N(t)=\left[\int\tilde{p}(t,\mathbf{x})\, d\mathbf{x}\right]\quad\mbox{(deterministic),}   \label{eq41}\\
&&N(t)=\left[\langle N(t,\cdot)\rangle \right] \quad\mbox{(stochastic), or}\label{eq42}\\ 
&& N(t)=\left[\int\left\langle\sum_{i=1}^{N(t,\cdot)}\delta_{\sigma_x}(\mathbf{x}-\mathbf{X}^i(t,\cdot))\right\rangle d\mathbf{x}\right]\!. \label{eq43}
\end{eqnarray}
Here $[x]$ is the integer part of the number $x$. Until about $t=30$ hours, the bulk of the marginal tip density pulse in Figures \ref{fig4} and \ref{fig5} has not reached the tumor at $x=L$. For all previous times, $N(t)$ is the same no matter whether it is calculated using the ensemble averages of (\ref{eq42}) or (\ref{eq43}). After this time, averaging directly counted number of tips, as in (\ref{eq42}), produces a higher number than the ensemble average density of (\ref{eq43}); see the inset of Figure \ref{fig6}. The deterministic prediction of (\ref{eq41}) is under these two lines. It is remarkable that the predictions based on integrals of marginal tip densities exhibit the same trend whether they are follow from stochastic or deterministic descriptions. The discrepancies are due to the fact that the {\em deterministic} pulse shown in Figure \ref{fig5} arrives {\em earlier} to the tumor than the {\em stochastic} pulse of Figure \ref{fig4} and, therefore, the deterministic marginal tip density is somewhat lower. Recall that the boundary condition \eqref{eq26} discards all tips that have arrived at the line $x=L$, but that some tips may not have arrived at the tumor ($|y|<b$). The leading front of the marginal density has a nonzero value at $x=L$, $|y|<b$ (the tumor) even if the actual vessel tips have not yet arrived there. This explains the discrepancies between the results of (\ref{eq42}) and (\ref{eq43}) shown in the inset of Figure \ref{fig6}. On the other hand, the agreement between the predictions based on stochastically or deterministically calculated marginal tip densities (except for the slightly faster deterministic pulse) shows that the deterministic description is a faithful approximation of the ensemble averaged stochastic description {\em provided the anastomosis coefficient is appropriately chosen}. 

\begin{figure}[ht]
\begin{center}
\includegraphics[width=12cm]{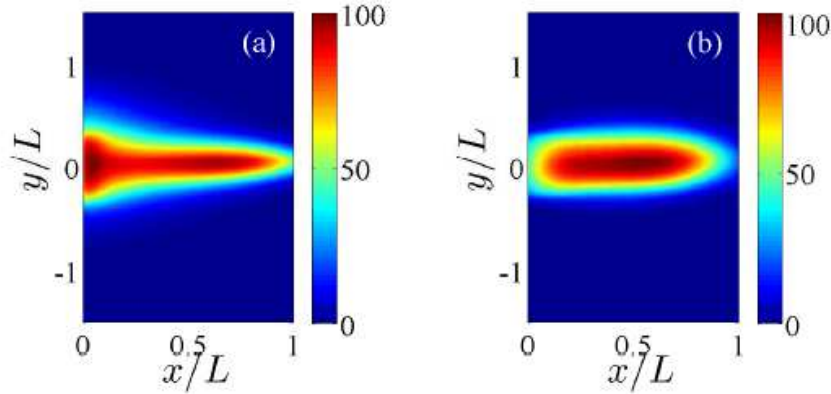} \qquad
\end{center}
\vskip -1cm
\caption{(Color online) Density plots of the overall network density $\int_0^t\tilde{p}(s,\mathbf{x})\, ds$ at $t=36$ hours calculated from (a) the deterministic description, and (b) the stochastic description. The total number of active tips are 61 and 66, respectively.
\label{fig7}}
\end{figure}

This is further shown by Figure \ref{fig7} that compares the final overall network density $\int_0^t\tilde{p}(s,\mathbf{x})\, ds$ as calculated from deterministic and stochastic descriptions. The flux of vessel tips injected from the primary vessel at $x=0$ produces a larger tip density there than is appreciated by ensemble averages of the stochastic process. This is also seen in Figure \ref{fig6}. There we observe that deterministic and stochastic descriptions predict a similar total number of tips until some of them begin to arrive at the tumor. It seems that the deterministic density is constrained to a narrower region by the chemotactic force than it is the case for the ensemble averaged density. As a consequence, the deterministic tip density loses more tips to tumor arrival than that for the stochastic process. This explains that the pulse of tip density travels faster than that given by ensemble averages at the later stage of angiogenesis, once tips begin arriving at the tumor. Note that there are tips that move outside the central region shown in Figure \ref{fig1}, issue less branches and may not arrive at the tumor. The overall deterministic vessel network becomes narrower and more elongated as shown in the left panel of Figure \ref{fig7}. 

\section{Conclusions}
\label{sec:conclusions}
We have solved numerically a simple stochastic model of tumor induced angiogenesis for many realizations (replicas of the system that differ in the initial condition). Numerically calculated velocity fluctuations do not decay even as the number of vessel tips increases. This shows that the stochastic model is not self-averaging and therefore we cannot use the law of large numbers to derive a deterministic description. However by re-examining the derivation given in \cite{bon14}, we conclude that the same deterministic description holds for vessel tip densities calculated by averaging over replicas. The deterministic description consists of a reaction-diffusion equation for the TAF concentration coupled to a Fokker-Planck type equation for the vessel tip density. The latter contains a birth term corresponding to tip branching and a death integral term corresponding to anastomosis or tip merging. The coefficient of the latter term has to be fitted by comparison with the stochastic description: optimal selection produces a good fit for the evolution of the total number of tips, provided fitting is carried out for intermediate times {\em after} an initial transient (about 8 hours) and {\em before} vessel tips begin arriving at the tumor (about 30 hours, see Fig.~\ref{fig8}). The reason for leaving out initial and final transients is that the phenomenological boundary conditions used in the deterministic description do not represent the stochastic description with sufficient accuracy. How to improve boundary conditions is an open problem.

Our work also has a general message elicited by the angiogenesis model: in stochastic models containing birth and death processes in addition to Brownian motion (Langevin equations), the death processes may preclude reaching the large number of individuals required to have self-averaging and a deterministic description based on the law of large numbers and the propagation of molecular chaos  for a single replica. Nevertheless, deterministic equations for macroscopic densities and fluxes may follow from the usual  law of large numbers applied to ensemble averages over a large number of replicas. 

\acknowledgments
This work has been supported by the Spanish Ministerio de Econom\'\i a y Competitividad grants
FIS2011-28838-C02-01 and MTM2014-56948-C2-2-P. VC  has been supported  by a Chair of Excellence UC3M-Santander at the Universidad Carlos III de Madrid. We thank Daniela Morale (University of Milan) for fruitful discussions and Mariano Alvaro (Universidad Carlos III) for letting us using his code for solving the system of deterministic equations.

\end{document}